\documentclass{mem}
\usepackage{natbib}
\usepackage{txfonts}
\usepackage{balance}
\usepackage{graphicx}
\usepackage{flushend}
\usepackage[a4paper,breaklinks,pdftex]{hyperref}
\idline{01}{001}
\begin{document}
\def\teff{$T\rm_{eff }$}
\def\kms{$\mathrm {km s}^{-1}$}

\title{
%New Observations on Planes of Satellite Galaxies
What new observations tell us about Planes of Satellite Galaxies
}

%   \subtitle{A Persistent Problem for Dark-Matter Based Cosmology}

\author{
Marcel \, S. \,Pawlowski\inst{1} 
          }

\institute{
Leibniz-Institute for Astrophysics,
An der Sternwarte 16, 
14482 Potsdam, 
Germany
\email{mpawlowski@aip.de}
}

\authorrunning{Pawlowski }

\titlerunning{New Observations of Planes of Satellites}

\date{Received: Day Month Year; Accepted: Day Month Year}

\abstract{
I briefly discuss the current state of the Planes of Satellite Galaxies Problem in light of some new observational data for satellite galaxies of the Milky Way, Andromeda, Centaurus A, and other systems beyond the Local Group. In particular, I present how a new proper motion measurment for Leo\,I enhances the overall orbital coherence among the MW's classical satellites and thus its tension with cosmological expectations.
}
\maketitle{}

\section{Introduction}

Cosmological simulations consistently predict a highly chaotic tangle of satellite dwarf galaxies around their hosts. Yet, the observed situation resembles more of an ordered choreography: satellite galaxies around the Milky Way (MW), Andromeda (M31), and Centaurus\,A (CenA) are aligned along so-called planes of satellite galaxies, flattened spatial arrangements of dwarf galaxies around their hosts \citep{PawlowskiKroupa2020, Ibata2013, Mueller2018}. Many satellites also appear to move along these structures in a common direction, either revealed by proper motions for the MW dwarfs, or hinted at by line-of-sight velocities for the structures beyond the MW.

Satellite systems with a similar coherence are exceedingly rare in cosmological simulations, with at most a few out of 1000 simulated systems approaching the degree of phase-space correlation found around the three best-observed host galaxies \citep{PawlowskiKroupa2020, Ibata2014, Mueller2018}. 
In contrast to other small-scale problems of cosmology that concern the internal properties of dwarf galaxies, the overall positions and motions of the dwarfs are not strongly affected by the details of baryonic physics prescriptions, and thus are a more robust test of dark-matter based cosmological models. Consequently, planes of satellite galaxies are considered to constitute a strong tension with cosmological expectations; \citealt{Sales2022, BoylanKolchin2021, Pawlowski2021NatAs}). 
%Planes of satellites thus pose a serious challenge, one that persistently lacks a clear avenue to a solution unless models beyond ΛCDM are considered.

While a lot of attention has been focussed on comparisons to cosmological simulations (e.g. \citealt{Samuel2021, Pham2022}), the inferred results are often affected more by the chosen methodology and metrics to quantify planarity and kinematic coherence of satellite structures than by actual differences among simulations. Putative disagreements in conclusions sometimes even appear to be mainly based on contradictory interpretations among different researchers, despite them reporting essentially the same degree of tension between the observed structures and the frequency of finding analogs of these in the simulations. It is then more a matter of the underlying research  philosophy whether a frequency of analogs $\lesssim 1:1000$\ constitutes a problem for our cosmological model (a limit difficult to exceed with current simulations as much larger high-resolution volumes would be required to increase the sample size of simulated hosts sufficiently to push to much higher significance).

True progress in understanding these satellite structures and their origins is therefore more likely to come from an improved understanding of the \textit{observed} structures. Recently, a number of new observational results have been published that in the following will be briefly reviewed and placed into context of the debate around planes of satellite galaxies as a challenge to the prevailing, dark matter based cosmological model.

\section{Recent Observational Progress}

%\begin{figure*}[t!]
\begin{figure*}[]
\center
\resizebox{1.0\hsize}{!}{\includegraphics[clip=true]{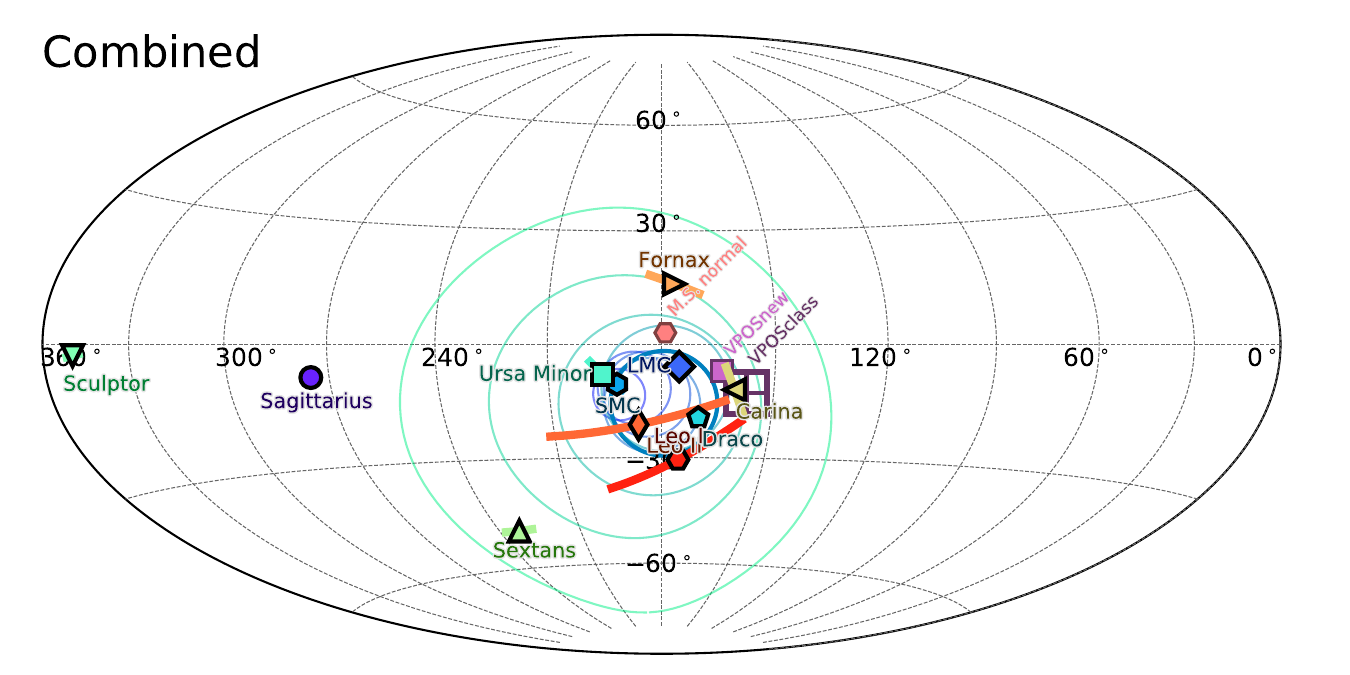}}
%\resizebox{\hsize}{!}{\includegraphics[clip=true]{ASP_Combined_NewLeoIpm.pdf}}
\caption{\footnotesize
Distribution of the orbital poles of the 11 classical MW satellites, adopting the combined proper motions from \citet{PawlowskiKroupa2020} for all but Leo\,I, for which the new multistudy average of \citet{CasettiDinescu2022} is adopted, which combines HST and Gaia data. The great-circle segments denote the orbital pole uncertainties, while the circles show the direction the spherical standard distance $\Delta_{\mathrm{std}}$, of the $k = [3, ..., 11]$\ most concentrated orbital poles. The circle for $k = 7$ is emphasized. 
%The directions of the normal vectors to the VPOS and to the Magellanic Stream are shown in purple.
}
\label{orbitalpoles}
\end{figure*}

\subsection{Milky Way's Vast Polar Structure}
%\subsection{Milky Way: a new proper motion for Leo\,I}

Among the 11 classical satellite galaxies of the MW, the most distant object Leo\,I has thus far been considered as likely not orbitally aligned with the overall Vast Polar Structure (VPOS) because existing Hubble Space Telescope (HST) and Gaia proper motions resulted in an orbital pole direction $>1\sigma$\ outside the clustering of orbital poles present for the majority of classical satellites. \citet{CasettiDinescu2022} have measured the absolute proper motion of Leo\,I using a 10-year baseline data from WFPC2 on HST in combination with Gaia EDR3. They find that the orbital pole of Leo\,I resulting from their new measurement is in good agreement with the direction of the VPOS normal.

How does this new proper motion change the overall situation of the orbital alignment of the classical satellites with the VPOS? Adopting their combined multistudy average of $(\mu_\alpha, \mu_\delta) = (-0.036 \pm 0.016, -0.130 \pm 0.010)\,\mathrm{mas\,yr}^{-1}$, I confirm that Leo\,I's orbital pole is closely aligned with the pronounced clustering of orbital poles (Fig. \ref{orbitalpoles}, Leo\,I in red).

\begin{figure}[]
\center
\resizebox{1.0\hsize}{!}{\includegraphics[clip=true]{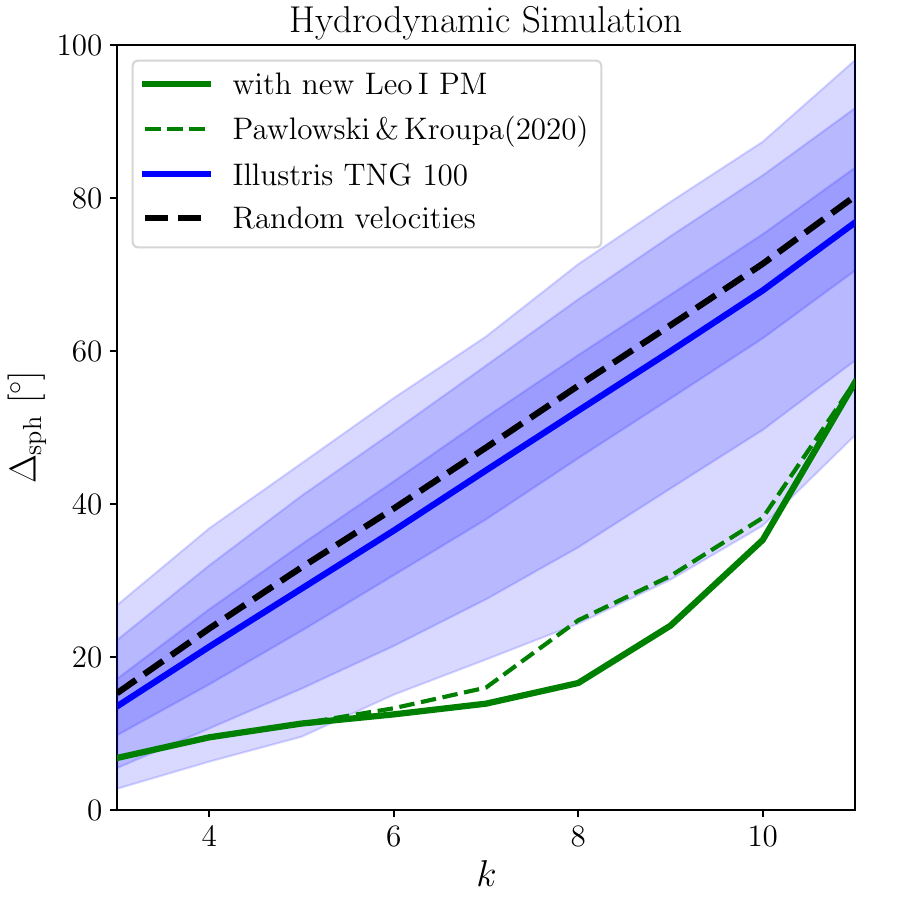}}
\caption{
\footnotesize
      Comparison of the spherical standard distances $\Delta_\mathrm{std}$\ of the $k$\ most-concentrated orbital poles with the expected distribution from the IllustrisTNG 100 simulation. The observed distribution is shown in green, comparing the proper motion sample combining Gaia and HST data \citep{PawlowskiKroupa2020} with the same after substituting Leo\,I's new proper motion. The new measurement clearly results in more compact distribution of orbital poles and thereby increased tension with cosmological expectations (contours contain 50, 90 and 99 per cent of all simulation's realizations).
%      The median $\Delta_\mathrm{std}$\ adopting random velocities for the satellites is shown as a black dashed line, demonstrating a light enhancement of orbital correlation among the simulated systems.
}
\label{simcomparison}
\end{figure}

Employing the orbital pole clustering measure of \citet{PawlowskiKroupa2020}, this new orbital pole results in a substantally tighter overall clustering (Fig. \ref{simcomparison}). For example, the $k=7$\ tightest clustered poles now have a spherical standard deviation of only $\Delta_{\mathrm{sph}}(k=7) = 13.9^\circ$\ (down from $16.0^\circ$), while the eight most tightly clustered ones now have $\Delta_{\mathrm{sph}}(k=8) = 16.6^\circ$\ (down from $24.8^\circ$). This is easily understood: with the new measurement, eight instead of seven of the classical satellites co-orbit.

In comparison to cosmological expectations, this increases the previously reported tension as already the previous orbital pole clustering was exteremely unlikely to be realized in MW analogs drawn from the IllustrisTNG-100 simulation \citep{Pillepich2018,PawlowskiKroupa2020}, and the new Leo\,I proper motion further increases this tension as demonstrated in Fig. \ref{simcomparison}.

\subsection{Great Plane of Andromeda}
%\subsection{Andromeda: first proper motions for two on-plane dwarf galaxies}

We observe M31's plane of satellites approximately edge-on. This allowed \citet{Ibata2013} to use line-of-sight velocities to test for kinematic coherence. They found that 13 out of the 15 on-plane satellites in their sample show a coherence indicative of a rotating satellite plane. In light of the behaviour of the classical MW dwarfs which predominantly co-orbit as inferred from their full 3D velocities, this suggests to interpret the line-of-sight correlation of the M31 system to also indicate a rotating structure. However, cosmological simulations suggest that if a satellite structure vaguely resembling that of M31 is found, many of it's satellites should quickly move off the plane.

In light of this, the first proper motion measurements for NGC\,147 and NGC\,185, two satellite plane members of the M31 system, by \citet{Sohn2020} provides a first test of the hypothesis that the satellite plane consists of co-orbiting satellites. They find both consistent with orbiting along the satellite plane, and to move in the direction expected from the line-of-sight velocity trend. 

However, while such behaviour is generally unexpected from cosmological simulations, \citet{PawlowskiSohn2021} have shown that two satellites alone do not suffice to firmly rule out the possibility that the structure is not orbitally coherent, since a few on-plane satellites can be expected to orbit along a putative structure by chance. More conclusive information will thus require measuring the proper motions of additional on-plane objects. Luckily, such measurements are under way and will soon increase the number of M31 satellites for which full 3D velocities are known.

\subsection{Centaurus\,A Satellite Plane}
%\subsection{Centaurus\,A: kinematic coherence confirmed}
\citet{Mueller2018} had used archival data to show that the known spatially flattened satellite distribution of CenA also displays a strong degree of kinematic coherence, and is highly unexpected for analog systems in cosmological simulations. Follow-up observations with MUSE provided additional line-of-sight velocities, which allowed \citet{Mueller2021} to test and confirm that the kinematic correlations remains as significant as previously reported, the tension with cosmological expectations remains also after employing comparisons of the new data to an updated hydrodynamical simulation, and the CenA satellite system thus remains problem for our prevailing dark matter based model of cosmology.

\subsection{Other systems}

Beyond observational progress in understanding the three best-studied cases of known satellite galaxy planes, some recent contributions have expanded such studies to additional systems. 
\citet{MartinezDelgado2021} report that NGC\,253 is surrounded by a flattened dwarf galaxy distribution, which might be connected to the presence of a cosmic filament and could constitute a satellite plane in the making. However, additional observations of distances \citep{MutluPakdil2022} and most importantly velocities, of the satellite galaxy candidates are needed to better understand the structure of the dwarf galaxy system and perform more meaningful comparisons to cosmological expectations.
\citet{Paudel2021} report that the galaxy NGC\,2750 is surrounded by dwarfs with pronounced kinematic correlation, emphasizing the systems similarity to that of CenA, though the number of observed dwarfs is considerably smaller at the moment and no detailed compaison to expectations has been perforemd yet.
\citet{Heesters2021} approach the issue in a statistical sense and present an analysis of dwarfs around 119 host galaxies from the MATLAS survey. While they only have access to projected positions and the sample will suffer some contamination by background galaxies, they do identify 31 flattened structures with high significance. This might indicate that satellite planes are indeed a persistent feature for a variety of hosts.

\section{Conclusions}

The phase-space distribution of satellite galaxies is a powerful test of cosmological predictions \citep{Pawlowski2021review}, in particular because it is largely unaffected by baryonic physics.
Recent observations of known satellite plane systems increases the orbital alignment of the classical MW satellites and strengthens the tension with cosmological expectations,
provide first evidence for orbital alignment in the M31 system that is generally unexpected for analogs in cosmological simulations, 
and confirm the tension posed by the CenA system with such expectations.
Furthermore, additional indications for similar satellite structures were found for external hosts.
Planes of satellite galaxies can thus be considered a persistent problem for standard cold dark matter based cosmology, motivating some researchers to follow alternative models in search of a solution. This includes tidal dwarf galaxy formation in a Modified Newtonian Dynamics framework (e.g. \citealt{Bilek2021}), or exotic dark matter models such as one where satellite galaxies might interact and align with domain walls in the dark sector \citep{Naik2022}. Further observational projects are ongoing and promise to provide additional insights into the nature and origin of these peculiar structures.

\begin{acknowledgements}
I acknowledge funding of a Leibniz-Junior Research Group (project number J94/2020) 
and a KT Boost Fund by the German Scholars Organization \& Klaus Tschira Stiftung.
\end{acknowledgements}

\bibliographystyle{aa}
\bibliography{refs_pawlowski}

\end{document}